\documentstyle[12pt]{article}

\begin{document}

\author{C. Bizdadea\thanks{%
e-mail address: bizdadea@hotmail.com}, E. M. Cioroianu and S. O. Saliu%
\thanks{%
e-mail addresses: osaliu@central.ucv.ro or odile\_saliu@hotmail.com} \\
Department of Physics, University of Craiova\\
13, A. I. Cuza Str., Craiova RO-1100, Romania}
\title{Irreducible Hamiltonian BRST approach to topologically coupled
abelian forms}
\maketitle

\begin{abstract}
An irreducible Hamiltonian BRST approach to topologically coupled $p$- and $%
\left( p+1\right) $-forms is developed. The irreducible setting is enforced
by means of constructing an irreducible Hamiltonian first-class model that
is equivalent from the BRST point of view to the original redundant theory.
The irreducible path integral can be brought to a manifestly Lorentz
covariant form.

\end{abstract}

\section{Introduction}

The typical feature of $p$-form gauge theories, namely, the reducibility
allows their link with string theory and supergravity models \cite{1}--\cite
{6}. Recently, $p$-form gauge theories have attracted attention in relation
with their characteristic cohomology \cite{7} and also with their
applications in higher dimensional bosonisation \cite{8a}. From the point of
view of the BRST quantization, theories involving $p$-forms implies the
introduction of ghost fields with ghost number greater that one (ghosts of
ghosts, etc.), and, in the meantime, of a pyramid of non-minimal variables 
\cite{8}--\cite{16}. Interacting $p$-forms were analyzed within the
reducible Hamiltonian BRST framework in \cite{17}, being obtained the ghost
and auxiliary field structures necessary at the antifield BRST quantization.

The main result of this paper consists in proving that it is possible to
quantize $p$-form gauge theories with topological coupling along an
irreducible Hamiltonian BRST procedure. Our method basically relies on
replacing the original redundant first-class model with an irreducible one,
and on further quantizing the resulting irreducible first-class system
accordingly the standard Hamiltonian BRST lines. The derivation of the
irreducible first-class theory relies on requiring that all the antighost
number one co-cycles from the Koszul-Tate homology identically vanish under
a convenient redefinition of the antighost number one antighosts while the
number of physical degrees of freedom is kept unchanged with respect to the
initial model. As a consequence of our analysis, the two theories are found
physically equivalent, which further allows (from the BRST point of view)
the substitution of the Hamiltonian BRST quantization of the reducible model
with that of the irreducible system. Initially we approach topologically
coupled abelian $p$- and $\left( p+1\right) $-forms described by a quadratic
action \cite{19} and then discuss the more general case of interacting
abelian forms with topological coupling, inferring an irreducible Lagrangian
formulation implied by our Hamiltonian approach that can be conveniently
applied to the interacting case. Although the idea of transforming a set of
reducible first-class constraints into an irreducible one is addressed in 
\cite{12}, \cite{20}, it has not been either developed or applied until now
to the irreducible quantization of this type of models.

The paper is organized in four sections. In Section 2 we focus on the
construction of an irreducible Hamiltonian first-class theory starting with
topologically coupled abelian $p$- and $\left( p+1\right) $-form gauge
fields described by a quadratic action within the homological context of the
Koszul-Tate differential, and provide the associated irreducible Hamiltonian
BRST symmetry. We then find by means of standard BRST Hamiltonian arguments
that it is permissible to replace the redundant Hamiltonian BRST symmetry
with the irreducible one, and infer the irreducible path integral with the
help of a suitable gauge-fixing fermion. Section 3 is devoted to the
extension of our irreducible procedure to the interacting case. There, we
work with a model of irreducible Hamiltonian first-class system and find
that the resulting Lagrangian gauge theory displays some manifestly Lorentz
covariant irreducible gauge transformations. The Lagrangian setting is
adequate for an irreducible approach to higher-order interacting gauge
theories with topological coupling. In Section 4 we expose the final
conclusions.

\section{Irreducible Hamiltonian BRST analysis}

In this section we construct the path integral for topologically coupled
abelian $p$- and $\left( p+1\right) $-form gauge fields in the context of an
irreducible Hamiltonian BRST procedure. In view of this, we perform the
canonical analysis of the starting quadratic Lagrangian action and observe
that this model is subject to some abelian first-class constraints that are $%
p$-stage reducible. The first-step of our irreducible approach consists in
the construction of an irreducible first-class set of constraints
corresponding to the initial redundant ones based on homological aspects.
This purpose is attained by means of making the original antighost number
one co-cycles from the reducible Koszul-Tate complex to vanish identically
under a proper redefinition of the antighost number one antighosts, and, in
the meantime, by maintaining the initial number of physical degrees of
freedom unchanged with respect to the irreducible background. The
implementation of these conditions yields an abelian irreducible first-class
constraint set, an associated first-class Hamiltonian and, moreover,
provides an irreducible Koszul-Tate complex corresponding to the original
reducible one. The construction is realized in a gradual manner starting
with the cases $p=1$ and $p=2$, and is further generalized to arbitrary
values of $p$. Next, we show that the irreducible BRST symmetry exists as it
satisfies the general grounds of homological perturbation theory. In the
sequel we investigate the correlation between the reducible and irreducible
Hamiltonian BRST symmetries and prove that the physical observables
underlying the reducible and irreducible theories coincide, which enables
the substitution of the Hamiltonian BRST quantization of the original model
with the BRST quantization of the irreducible system. Finally, we realize
the Hamiltonian BRST quantization of the irreducible model by using an
appropriate gauge-fixing fermion and non-minimal sector, inferring the
irreducible path integral, which is local and manifestly Lorentz covariant.

\subsection{Canonical analysis of the reducible model}

We start with the quadratic Lagrangian action%
\begin{eqnarray}\label{t1}
& &\tilde S_0^L\left[ A^{\mu _1\ldots \mu _p},H^{\mu _1\ldots \mu
_{p+1}}\right] =\int d^{2p+2}x\left( -\frac 1{2\cdot \left( p+1\right)
!}F_{\mu _1\ldots \mu _{p+1}}^2-\right. \nonumber \\ 
& &\left. \frac 1{2\cdot \left( p+2\right) !}F_{\mu _1\ldots \mu
_{p+2}}^2+\frac M{p!\cdot \left( p+2\right) !}\varepsilon _{\mu _1\ldots \mu
_{2p+2}}F^{\mu _1\ldots \mu _{p+2}}A^{\mu _{p+3}\ldots \mu _{2p+2}}\right) ,
\end{eqnarray}
where $\left( F_{\mu _1\ldots \mu _{p+1}},F_{\mu _1\ldots \mu _{p+2}}\right) 
$ stand for the field strengths respectively corresponding to the
antisymmetric tensor fields $\left( A^{\mu _1\ldots \mu _p},H^{\mu _1\ldots
\mu _{p+1}}\right) $, and $\varepsilon _{\mu _1\ldots \mu _{2p+2}}$ denote
the completely antisymmetric symbol in $\left( 2p+2\right) $ dimensions. The
notation $F_{\mu _1\ldots \mu _{p+1}}^2$ signifies $F_{\mu _1\ldots \mu
_{p+1}}F^{\mu _1\ldots \mu _{p+1}}$, and the same for the other square. It
is worthnote that the topological coupling present in the third term from
the right-hand side of (\ref{t1}) is a generalization of the Chern-Simons
coupling introduced by Jackiw, Deser, {\it et al} \cite{20a}--\cite{20c}.

From the canonical approach of (\ref{t1}), one infers the first-class
constraints 
\begin{equation}
\label{t2}\tilde \gamma _{i_1\ldots i_{p-1}}^{(1)}\equiv \pi _{0i_1\ldots
i_{p-1}}\approx 0, 
\end{equation}
\begin{equation}
\label{t3}\bar \gamma _{i_1\ldots i_p}^{(1)}\equiv \Pi _{0i_1\ldots
i_p}\approx 0, 
\end{equation}
\begin{eqnarray}\label{t4}
& &\tilde G_{i_1\ldots i_{p-1}}^{(2)}\equiv -p\partial ^i\pi _{ii_1\ldots
i_{p-1}}+\nonumber \\ 
& &\left( -\right) ^{p+1}\frac M{\left( p-1\right) !\cdot \left(
p+2\right) !}\varepsilon _{0i_1\ldots i_{p-1}j_1\ldots j_{p+2}}F^{j_1\ldots
j_{p+2}}\approx 0, 
\end{eqnarray}
\begin{equation}
\label{t5}\bar G_{i_1\ldots i_p}^{(2)}\equiv -\left( p+1\right) \partial
^i\Pi _{ii_1\ldots i_p}\approx 0, 
\end{equation}
and the canonical Hamiltonian%
\begin{eqnarray}\label{t6}
& &\tilde H=
\int d^{2p+1}x\left( -\frac{p!}2\pi _{i_1\ldots i_p}\pi ^{i_1\ldots
i_p}-\frac{\left( p+1\right) !}2\Pi _{i_1\ldots i_{p+1}}\Pi ^{i_1\ldots
i_{p+1}}+\right. \nonumber \\ 
& &\frac M{p!}\varepsilon _{0i_1\ldots i_{p+1}j_1\ldots j_p}\Pi ^{i_1\ldots
i_{p+1}}A^{j_1\ldots j_p}+\frac 1{2\cdot \left( p+1\right) !}F_{i_1\ldots
i_{p+1}}F^{i_1\ldots i_{p+1}}+\nonumber \\ 
& &\frac 1{2\cdot \left( p+2\right) !}F_{i_1\ldots i_{p+2}}F^{i_1\ldots
i_{p+2}}+\frac{M^2}{2\cdot p!}A_{i_1\ldots i_p}A^{i_1\ldots i_p}+\nonumber \\ 
& &\left. A^{0i_1\ldots i_{p-1}}\tilde G_{i_1\ldots
i_{p-1}}^{(2)}+H^{0i_1\ldots i_p}\bar G_{i_1\ldots i_p}^{(2)}\right) . 
\end{eqnarray}
The secondary constraints (\ref{t4}) and (\ref{t5}) are $\left( p-1\right) $%
, respectively, $p$-stage reducible, with the reducibility relations given
by 
\begin{equation}
\label{t7}Z_{\;\;j_1\ldots j_{p-2}}^{i_1\ldots i_{p-1}}\tilde G_{i_1\ldots
i_{p-1}}^{(2)}=0,\;\bar Z_{\;\;j_1\ldots j_{p-1}}^{i_1\ldots i_p}\bar
G_{i_1\ldots i_p}^{(2)}=0, 
\end{equation}
\begin{equation}
\label{t8}Z_{\;\;j_1\ldots j_{p-k-1}}^{i_1\ldots i_{p-k}}Z_{\;\;l_1\ldots
l_{p-k-2}}^{j_1\ldots j_{p-k-1}}=0,\;k=1,\ldots ,p-2, 
\end{equation}
\begin{equation}
\label{t11}\bar Z_{\;\;j_1\ldots j_{p-k}}^{i_1\ldots i_{p-k+1}}\bar
Z_{\;\;l_1\ldots l_{p-k-1}}^{j_1\ldots j_{p-k}}=0,\;k=1,\ldots ,p-1, 
\end{equation}
and the $k$th order reducibility functions expressed by 
\begin{equation}
\label{t9}Z_{\;\;j_1\ldots j_{p-k-1}}^{i_1\ldots i_{p-k}}=\frac 1{\left(
p-k-1\right) !}\partial _{}^{\left[ i_1\right. }\delta
_{\;\;j_1}^{i_2}\ldots \delta _{\;\;j_{p-k-1}}^{\left. i_{p-k}\right]
},\;k=1,\ldots ,p-1, 
\end{equation}
\begin{equation}
\label{t12}\bar Z_{\;\;j_1\ldots j_{p-k}}^{i_1\ldots i_{p-k+1}}=\frac
1{\left( p-k\right) !}\partial _{}^{\left[ i_1\right. }\delta
_{\;\;j_1}^{i_2}\ldots \delta _{\;\;j_{p-k}}^{\left. i_{p-k+1}\right]
},\;k=1,\ldots ,p. 
\end{equation}
The notations $\pi _{0i_1\ldots i_{p-1}}$ and $\pi _{i_1\ldots i_p}$ signify
the canonical momenta conjugated with the corresponding $A$'s, while $\Pi
_{0i_1\ldots i_p}$ and $\Pi _{i_1\ldots i_{p+1}}$ stand for the canonical
momenta associated with the $H$'s. The notation $\left[ i_1\ldots
i_{p-k}\right] $ signifies antisymmetry with respect to the indices between
brackets. In the sequel we work with the conventions $f^{i_1\ldots i_m}=f$
if $m=0$, and $f^{i_1\ldots i_m}=0$ if $m<0$.

\subsection{Construction of irreducible constraints}

Initially, we obtain an irreducible model corresponding to topologically
coupled abelian $p$- and $\left( p+1\right) $-form gauge fields by means of
homological arguments and by requesting the preservation of the number of
physical degrees of freedom with respect to the redundant model. In this
context, we derive an irreducible first-class set associated with the
reducible constraints (\ref{t4}--\ref{t5}). In order to clarify the main
aspects linked to our irreducible treatment, we gradually investigate the
cases $p=1$ and $p=2$, and then generalize the construction to an arbitrary $%
p$.

\subsubsection{The case $p=1$}

The constraints (\ref{t4}--\ref{t5}) take in this situation the concrete
form 
\begin{equation}
\label{tp1}\tilde G^{(2)}\equiv -\partial ^j\pi _j+\frac M6\varepsilon
_{0jkl}F^{jkl}\approx 0, 
\end{equation}
\begin{equation}
\label{tp2}\bar G_i^{(2)}\equiv -2\partial ^j\Pi _{ji}\approx 0, 
\end{equation}
and are first-stage reducible, the reducibility relations being expressed by 
\begin{equation}
\label{tp3}Z^i\bar G_i^{(2)}\equiv \partial ^i\bar G_i^{(2)}=0. 
\end{equation}
The reducible Hamiltonian BRST symmetry $s_R=\delta _R+D_R+\cdots $ involves
two crucial graded differentials. One of them ($\delta _R$) is called the
Koszul-Tate differential and realizes an homological resolution of smooth
functions defined on the first-class constraint surface. Its graduation is
governed by the antighost number ($antigh$), and we have that $antigh\left(
\delta _R\right) =-1$. The main property of $\delta _R$ is the acyclicity at
non-vanishing antighost numbers. The other one ($D_R$) is known as a model
of exterior derivative along the gauge orbits and accounts for the gauge
invariances implied by the presence of the first-class constraints. The
degree of $D_R$ is named pure ghost number ($pure\;gh$), and is defined like 
$pure\;gh\left( D_R\right) =1$. In the case $p=1$ the reducible Koszul-Tate
complex includes the antighost number one fermionic antighosts ${\cal P}_2$
and $P_{2i}$, being defined through the relations 
\begin{equation}
\label{tp4}\delta _Rz^A=0, 
\end{equation}
\begin{equation}
\label{tp5}\delta _R{\cal P}_2=-\tilde G^{(2)}, 
\end{equation}
\begin{equation}
\label{tp6}\delta _RP_{2i}=-\bar G_i^{(2)}, 
\end{equation}
where $z^A$ is any of the fields $A^\mu $, $H^{\mu \nu }$ or their momenta.
With the help of the definitions (\ref{tp6}) and the reducibility relations (%
\ref{tp3}), it follows that there appear a non trivial co-cycle in the
homology of $\delta _R$, of the type 
\begin{equation}
\label{tp7}\bar \mu =\partial ^iP_{2i}. 
\end{equation}
In order to restore the $\delta _R$-exactness of this co-cycle and thus the
acyclicity of the Koszul-Tate differential it is necessary to enhance the
antighost spectrum with the antighost number two bosonic antighost $\bar
\lambda $ and to set 
\begin{equation}
\label{tp8}\delta _R\bar \lambda =-\partial ^iP_{2i}. 
\end{equation}
The idea of transforming this reducible model into an irreducible one is
based on redefining the antighost number one antighosts $P_{2i}$ involved
with the co-cycle (\ref{tp7}) like 
\begin{equation}
\label{tp9}P_{2i}\rightarrow P_{2i}^{\prime }=D_{\;\;i}^jP_{2j}, 
\end{equation}
such that the new co-cycle of the type (\ref{tp7}), namely, 
\begin{equation}
\label{tp10}\bar \mu ^{\prime }=\partial ^iP_{2i}^{\prime } 
\end{equation}
vanishes identically. As a consequence, the new co-cycle $\bar \mu ^{\prime
} $ is trivial without adding the antighost number two antighost $\bar
\lambda $, hence the resulting model is irreducible. In view of this we
choose the matrix $D_{\;\;i}^j$ to satisfy the properties 
\begin{equation}
\label{tp11}\partial ^iD_{\;\;i}^j=0, 
\end{equation}
\begin{equation}
\label{tp12}D_{\;\;i}^j\bar G_j^{(2)}=\bar G_i^{(2)}. 
\end{equation}
Taking into account (\ref{tp6}), (\ref{tp9}) and (\ref{tp12}), we have that 
\begin{equation}
\label{tp13}\delta P_{2i}^{\prime }=-\bar G_i^{(2)}, 
\end{equation}
while the properties (\ref{tp11}) yield that $\bar \mu ^{\prime }$ is indeed
vanishing 
\begin{equation}
\label{tp14}\bar \mu ^{\prime }\equiv 0. 
\end{equation}
In (\ref{tp13}) we used the notation $\delta $ instead of $\delta _R$ in
order to emphasize the irreducibility of the new approach. Thus, if the
equations (\ref{tp11}--\ref{tp12}) possess solutions, then the co-cycle $%
\bar \mu ^{\prime }$ vanishes identically and the theory becomes
irreducible, the presence of the antighost $\bar \lambda $ being useless.
The solution to the equations (\ref{tp11}--\ref{tp12}) exists and is given
by 
\begin{equation}
\label{tp15}D_{\;\;i}^j=\delta _{\;\;i}^j-\frac{\partial ^j\partial _i}%
\Delta , 
\end{equation}
where $\Delta =\partial ^k\partial _k$. Replacing (\ref{tp15}) in (\ref{tp13}%
) we arrive at 
\begin{equation}
\label{tp16}\delta P_{2i}-\frac{\partial _i}\Delta \delta \left( \partial
^jP_{2j}\right) =-\bar G_i^{(2)}. 
\end{equation}
The relations (\ref{tp16}) describe the action of the Koszul-Tate
differential underlying an irreducible model. At this point we explore the
request on the equality between the numbers of physical degrees of freedom
associated with the reducible and irreducible theories. The original
reducible theory has three physical degrees of freedom, while the
irreducible theory possesses two physical degrees of freedom as the set (\ref
{tp2}) will be replaced by a corresponding set of three independent
first-class constraints. This is why we need to supplement the original
field/momentum spectrum of the irreducible theory with an extra canonical
bosonic pair, to be denoted by $\left( H,\Pi \right) $. With these
supplementary variables at hand, the number of physical degrees of freedom
associated with the irreducible model is now equal to three. We demand that $%
\Pi $ is the non vanishing solution to the equation 
\begin{equation}
\label{tp17}\delta \left( \partial ^iP_{2i}\right) =\triangle \Pi . 
\end{equation}
The last condition together with the invertibility of $\triangle $ guarantee
the irreducibility of the new theory because the last equation possesses
non-vanishing solutions if and only if $\delta \left( \partial
^iP_{2i}\right) \neq 0$, hence if and only if (\ref{tp7}) is not a co-cycle.
Inserting (\ref{tp17}) in (\ref{tp16}) we infer that 
\begin{equation}
\label{tp18}\delta P_{2i}=-\bar G_i^{(2)}+\partial _i\Pi \equiv -\bar \gamma
_i^{(2)}, 
\end{equation}
which signify the definitions of $\delta $ on the antighost number one
antighosts associated with an irreducible model possessing the irreducible
constraints 
\begin{equation}
\label{tp19}\bar \gamma _i^{(2)}\equiv \bar G_i^{(2)}-\partial _i\Pi \approx
0, 
\end{equation}
instead of the reducible constraints (\ref{tp2}) of the original theory. In
conclusion, we constructed an irreducible first-class constraint set
corresponding to topologically coupled abelian one- and two-form gauge
fields, of the type 
\begin{equation}
\label{tp20}\tilde \gamma ^{(1)}\equiv \pi _0\approx 0,\;\bar \gamma
_i^{(1)}\equiv \Pi _{0i}\approx 0, 
\end{equation}
\begin{equation}
\label{tp21}\tilde \gamma ^{(2)}\equiv -\partial ^j\pi _j+\frac
M6\varepsilon _{0jkl}F^{jkl}\approx 0,\;\bar \gamma _i^{(2)}\equiv
-2\partial ^j\Pi _{ji}-\partial _i\Pi \approx 0. 
\end{equation}

\subsubsection{The case $p=2$}

The constraints (\ref{t4}--\ref{t5}) are given in this case by 
\begin{equation}
\label{tp22}\tilde G_i^{(2)}\equiv -2\partial ^j\pi _{ji}-\frac
M{24}\varepsilon _{0ijklm}F^{jklm}\approx 0, 
\end{equation}
\begin{equation}
\label{tp23}\bar G_{ij}^{(2)}\equiv -3\partial ^k\Pi _{kij}\approx 0, 
\end{equation}
and are second-stage reducible, the first-stage reducibility relations being
given by 
\begin{equation}
\label{tp24}\partial ^i\tilde G_i^{(2)}=0,\;2\partial ^i\bar G_{ij}^{(2)}=0. 
\end{equation}
By introducing the fermionic antighosts ${\cal P}_{2i}$ and $P_{2ij}$ of
antighost number one, the Koszul-Tate operator acts like 
\begin{equation}
\label{tp25}\delta _R{\cal P}_{2i}=-\tilde G_i^{(2)}, 
\end{equation}
\begin{equation}
\label{tp26}\delta _RP_{2ij}=-\bar G_{ij}^{(2)}, 
\end{equation}
while its action on the original fields/momenta is vanishing. The
reducibility relations (\ref{tp24}) yield the antighost number one non
trivial co-cycles 
\begin{equation}
\label{tp27}\tilde \nu \equiv \partial ^i{\cal P}_{2i}, 
\end{equation}
\begin{equation}
\label{tp28}\bar \nu _j\equiv 2\partial ^iP_{2ij}. 
\end{equation}
In order to restore the acyclicity of $\delta _R$ we add the bosonic
antighost number two antighosts $\tilde \lambda $ and $\bar \lambda _i$, and
put 
\begin{equation}
\label{tp29}\delta _R\tilde \lambda =-\partial ^i{\cal P}_{2i}, 
\end{equation}
\begin{equation}
\label{tp30}\delta _R\bar \lambda _j=-2\partial ^iP_{2ij}. 
\end{equation}
Because of the second-stage reducibility relation, there appear a
supplementary non trivial co-cycle at antighost number two 
\begin{equation}
\label{tp31}\bar \nu \equiv \partial ^i\bar \lambda _i, 
\end{equation}
which is `killed' by means of introducing the fermionic antighost number
three antighost $\bar \lambda $ through 
\begin{equation}
\label{tp31a}\delta _R\bar \lambda =-\partial ^i\bar \lambda _i. 
\end{equation}

The passing to the irreducible model goes along the line employed at the
case $p=1$, namely, we enforce that the objects $\tilde \nu $ and $\bar \nu
_j$ are not closed in terms of the irreducible Koszul-Tate differential $%
\delta $, therefore not co-cycles. This request can be satisfied by adding
the bosonic canonical pairs $\left( A,\pi \right) $, $\left( H^i,\Pi
_i\right) $ whose momenta are the non vanishing solutions to the equations 
\begin{equation}
\label{tp33}\delta \left( \partial ^i{\cal P}_{2i}\right) =\triangle \pi , 
\end{equation}
\begin{equation}
\label{tp32}\delta \left( 2\partial ^iP_{2ij}\right) =\triangle \Pi _j. 
\end{equation}
Applying $\partial ^j$ on (\ref{tp32}) it follows that $\triangle \left(
\partial ^j\Pi _j\right) =0$, which further leads to 
\begin{equation}
\label{tp34}\partial ^j\Pi _j=0, 
\end{equation}
on account of the invertibility of $\triangle $. The prior relation is
nothing but a new constraint of the irreducible theory 
\begin{equation}
\label{tp35}\bar \gamma ^{(2)}\equiv -\partial ^j\Pi _j\approx 0, 
\end{equation}
which is necessary in order to maintain the number of physical degrees of
freedom for the irreducible model equal with that of the redundant theory.
Indeed, the number of independent constraints (\ref{tp22}--\ref{tp23}) is
equal to ten, hence the reducible model displays ten physical degrees of
freedom. The irreducible model will possess thirty independent constraint
functions corresponding to the reducible set (\ref{tp22}--\ref{tp23}) plus
the supplementary pairs $\left( A,\pi \right) $, $\left( H^i,\Pi _i\right) $%
, which gives eleven physical degrees of freedom. It is precisely the
presence of the new first-class constraint (\ref{tp33}) that restores the
number of physical degrees of freedom associated with the irreducible theory
to ten. We notice that the constraint function $\bar \gamma ^{(2)}$ is
irreducible with respect to (\ref{tp22}--\ref{tp23}), such that it does not
induce further antighost number one co-cycles. By introducing its antighost $%
P_2$ (which is fermionic of antighost number one), the corresponding action
of the irreducible Koszul-Tate operator reads as 
\begin{equation}
\label{tp36}\delta P_2=-\bar \gamma ^{(2)}. 
\end{equation}

Next, we perform the redefinition of the antighosts ${\cal P}_{2i}$ and $%
P_{2ij}$ in such a way that the new co-cycles of the type (\ref{tp27}--\ref
{tp28}) identically vanish. In this light, we remark that the constraint
functions in (\ref{tp22}--\ref{tp23}) are separately reducible, such that
the redefinition of the antighosts ${\cal P}_{2i}$ and $P_{2ij}$ can be done
in a way that does not mix these fields, namely, 
\begin{equation}
\label{tp37}{\cal P}_{2i}\rightarrow {\cal P}_{2i}^{\prime }=D_{\;\;i}^j%
{\cal P}_{2j}, 
\end{equation}
\begin{equation}
\label{tp38}P_{2ij}\rightarrow P_{2ij}^{\prime }=D_{\;\;ij}^{kl}P_{2kl}. 
\end{equation}
We demand that the matrices $D_{\;\;i}^j$ and $D_{\;\;ij}^{kl}$ are subject
to the conditions 
\begin{equation}
\label{tp39}\partial ^iD_{\;\;i}^j=0,\;2\partial ^iD_{\;\;ij}^{kl}=0, 
\end{equation}
\begin{equation}
\label{tp40}D_{\;\;i}^j\tilde G_j^{(2)}=\tilde
G_i^{(2)},\;D_{\;\;ij}^{kl}\bar G_{kl}^{(2)}=\bar G_{ij}^{(2)}. 
\end{equation}
On the one hand, with the help of the conditions (\ref{tp40}) and using (\ref
{tp37}--\ref{tp38}) we find that 
\begin{equation}
\label{tp41}\delta {\cal P}_{2i}^{\prime }=-\tilde G_i^{(2)}, 
\end{equation}
\begin{equation}
\label{tp42}\delta P_{2ij}^{\prime }=-\bar G_{ij}^{(2)}, 
\end{equation}
while, on the other hand, the properties (\ref{tp39}) yield that the new
co-cycles of the type (\ref{tp27}--\ref{tp28}) vanish identically, i.e., 
\begin{equation}
\label{tp43}\partial ^i{\cal P}_{2i}^{\prime }\equiv 0, 
\end{equation}
\begin{equation}
\label{tp44}2\partial ^iP_{2ij}^{\prime }\equiv 0. 
\end{equation}
The solution to the equations (\ref{tp39}--\ref{tp40}) exists and is
expressed by (\ref{tp15}) for $D_{\;\;i}^j$ and by 
\begin{equation}
\label{tp45}D_{\;\;ij}^{kl}=\frac 12\left( \delta _{\;\;i}^{\left[ k\right.
}\delta _{\;\;j}^{\left. l\right] }-\frac 1{\triangle }\delta
_{\;\;m}^{\left[ l\right. }\partial _{}^{\left. k\right] }\delta
_{\;\;\left[ j\right. }^m\partial _{\left. i\right] }^{}\right) . 
\end{equation}
Substituting the solutions (\ref{tp15}) and (\ref{tp45}) in the relations (%
\ref{tp41}--\ref{tp42}) and recalling that $\left( \pi ,\Pi _i\right) $ are
the non vanishing solutions to the equations (\ref{tp33}--\ref{tp32}), we
obtain 
\begin{equation}
\label{tp46}\delta {\cal P}_{2i}=-\tilde G_i^{(2)}+\partial _i\pi \equiv
-\tilde \gamma _i^{(2)}, 
\end{equation}
\begin{equation}
\label{tp47}\delta P_{2ij}=-\bar G_{ij}^{(2)}+\frac 12\partial _{\left[
i\right. }\Pi _{\left. j\right] }\equiv -\bar \gamma _{ij}^{(2)}, 
\end{equation}
which emphasize the irreducible constraints deriving from the reducible set (%
\ref{tp22}--\ref{tp23}) under the form 
\begin{equation}
\label{tp48}\tilde \gamma _i^{(2)}\equiv \tilde G_i^{(2)}-\partial _i\pi
\approx 0, 
\end{equation}
\begin{equation}
\label{tp49}\bar \gamma _{ij}^{(2)}\equiv \bar G_{ij}^{(2)}-\frac 12\partial
_{\left[ i\right. }\Pi _{\left. j\right] }\approx 0. 
\end{equation}
In conclusion, the irreducible model attached to two- and three-forms with
topological coupling is pictured by the irreducible first-class constraint
set 
\begin{equation}
\label{tp50}\tilde \gamma _i^{(1)}\equiv \pi _{0i}\approx 0,\;\bar \gamma
_{ij}^{(1)}\equiv \Pi _{0ij}\approx 0, 
\end{equation}
\begin{equation}
\label{tp51}\tilde \gamma _i^{(2)}\equiv -2\partial ^j\pi _{ji}-\frac
M{24}\varepsilon _{0ijklm}F^{jklm}-\partial _i\pi \approx 0, 
\end{equation}
\begin{equation}
\label{tp52}\bar \gamma _{ij}^{(2)}\equiv -3\partial ^k\Pi _{kij}-\frac
12\partial _{\left[ i\right. }\Pi _{\left. j\right] }\approx 0, 
\end{equation}
\begin{equation}
\label{tp53}\bar \gamma ^{(2)}\equiv -\partial ^j\Pi _j\approx 0. 
\end{equation}

\subsubsection{Generalization to arbitrary $p$}

Now, we are in the position to generalize the irreducible construction to
arbitrary values of $p$. The first step resides in deriving a reducible
theory involving more fields. To this end, we introduce the antisymmetric
bosonic canonical pairs 
\begin{equation}
\label{t13}\left( A^{j_1\ldots j_{p-2k-2}},\pi _{j_1\ldots
j_{p-2k-2}}\right) _{k=0,\cdots ,c},\;\left( H^{i_1\ldots i_{p-2k-1}},\Pi
_{i_1\ldots i_{p-2k-1}}\right) _{k=0,\cdots ,a}, 
\end{equation}
and, acting accordingly some homological arguments similar to those used
previously, we infer the following irreducible first-class set corresponding
to (\ref{t4}--\ref{t5}) 
\begin{equation}
\label{t26}\tilde \gamma _{i_1\ldots i_{p-2k-1}}^{(2)}\approx 0,\;k=0,\ldots
,a, 
\end{equation}
\begin{eqnarray}\label{t28}
& &\bar \gamma _{i_1\ldots i_{p-2k}}^{(2)}\equiv -
\left( p-2k+1\right) \partial
^i\Pi _{ii_1\ldots i_{p-2k}}-\nonumber \\
& &\partial _{\left[ i_1\right. }^{}\Pi _{\left. i_2\ldots
i_{p-2k}\right] }\approx 0,\;k=0,\ldots ,b,
\end{eqnarray}
with 
\begin{equation}
\label{t29}\tilde \gamma _{i_1\ldots i_{p-2k-1}}^{(2)}\equiv \left\{ 
\begin{array}{l}
\tilde G_{i_1\ldots i_{p-1}}^{(2)}-\partial _{\left[ i_1\right. }^{}\pi
_{\left. i_2\ldots i_{p-1}\right] },\;k=0, \\ 
-\left( p-2k\right) \partial ^i\pi _{ii_1\ldots i_{p-2k-1}}-\partial
_{\left[ i_1\right. }^{}\pi _{\left. i_2\ldots i_{p-2k-1}\right]
},\;k=1,\ldots ,a, 
\end{array}
\right. 
\end{equation}
where we employed the notations 
$$
a=\left\{ 
\begin{array}{c}
\frac p2-1,\; 
{\rm if}\;p\;{\rm even}, \\ \frac{p-1}2,\;{\rm if}\;p\;{\rm odd}, 
\end{array}
\right. ,b=\left\{ 
\begin{array}{c}
\frac p2,\; 
{\rm if}\;p\;{\rm even}, \\ \frac{p-1}2,\;{\rm if}\;p\;{\rm odd}, 
\end{array}
\right. ,c=\left\{ 
\begin{array}{c}
\frac p2-1,\; 
{\rm if}\;p\;{\rm even}, \\ \frac{p-3}2,\;{\rm if}\;p\;{\rm odd}. 
\end{array}
\right. 
$$

In order to infer a manifestly covariant path integral for the irreducible
theory it is still necessary to add some supplementary canonical pairs
subject to some additional constraints such that on the one hand the entire
set of resulting constraints is first-class and irreducible, and, on the
other hand, the number of physical degrees of freedom of the irreducible
theory remains unchanged as compared to that of the redundant model. First,
we introduce the antisymmetric bosonic canonical pairs 
\begin{equation}
\label{t14}\left( A^{0i_1\ldots i_{p-2k-3}},\pi _{0i_1\ldots
i_{p-2k-3}}\right) _{k=0,\cdots ,d},\;\left( H^{0i_1\ldots i_{p-2k-2}},\Pi
_{0i_1\ldots i_{p-2k-2}}\right) _{k=0,\cdots ,c}, 
\end{equation}
subject to the constraints 
\begin{equation}
\label{pi}\pi _{0i_1\ldots i_{p-2k-3}}\approx 0,\;\Pi _{0i_1\ldots
i_{p-2k-2}}\approx 0, 
\end{equation}
where $d$ is defined by%
$$
d=\left\{ 
\begin{array}{c}
\frac p2-2,\; 
{\rm if}\;p\;{\rm even}, \\ \frac{p-3}2,\;{\rm if}\;p\;{\rm odd}. 
\end{array}
\right. 
$$
We redenote the constraints (\ref{t2}--\ref{t3}) together with (\ref{pi}) by 
\begin{equation}
\label{t25}\tilde \gamma _{i_1\ldots i_{p-2k-1}}^{(1)}\equiv \pi
_{0i_1\ldots i_{p-2k-1}}\approx 0,\;k=0,\cdots ,a, 
\end{equation}
\begin{equation}
\label{t27}\bar \gamma _{i_1\ldots i_{p-2k}}^{(1)}\equiv \Pi _{0i_1\ldots
i_{p-2k}}\approx 0,\;k=0,\cdots ,b. 
\end{equation}
Thus, the irreducible model is described until now by the irreducible
abelian first-class constraints (\ref{t26}--\ref{t28}) and (\ref{t25}--\ref
{t27}). We take the first-class Hamiltonian with respect to these
constraints under the form%
\begin{eqnarray}\label{t30}
& &\tilde H^{\prime }=\int d^{2p+1}x\left( -\frac{p!}2\pi _{i_1\ldots i_p}\pi
^{i_1\ldots i_p}-\frac{\left( p+1\right) !}2\Pi _{i_1\ldots i_{p+1}}\Pi
^{i_1\ldots i_{p+1}}+\right. \nonumber \\ 
& &\frac M{p!}\varepsilon _{0i_1\ldots i_{p+1}j_1\ldots j_p}\Pi ^{i_1\ldots
i_{p+1}}A^{j_1\ldots j_p}+\frac 1{2\cdot \left( p+1\right) !}F_{i_1\ldots
i_{p+1}}F^{i_1\ldots i_{p+1}}+\nonumber \\ 
& &\frac 1{2\cdot \left( p+2\right) !}F_{i_1\ldots i_{p+2}}F^{i_1\ldots
i_{p+2}}+\frac{M^2}{2\cdot p!}A_{i_1\ldots i_p}A^{i_1\ldots i_p}+ 
\nonumber \\
& &\left. \sum_{k=0}^aA^{0i_1\ldots i_{p-2k-1}}\tilde \gamma
_{i_1\ldots i_{p-2k-1}}^{(2)}+\sum_{k=0}^bH^{0i_1\ldots i_{p-2k}}\bar \gamma
_{i_1\ldots i_{p-2k}}^{(2)}\right) .
\end{eqnarray}
Second, to every pair (\ref{t13}) we associate two more antisymmetric
bosonic pairs, respectively denoted by 
\begin{equation}
\label{t31}\left( B^{(1)i_1\cdots i_{p-2k-2}},\pi _{i_1\cdots
i_{p-2k-2}}^{(1)}\right) ,\;\left( B^{(2)i_1\cdots i_{p-2k-2}},\pi
_{i_1\cdots i_{p-2k-2}}^{(2)}\right) ,\;k=0,\cdots ,c, 
\end{equation}
\begin{equation}
\label{t32}\left( V^{(1)i_1\cdots i_{p-2k-1}},\Pi _{i_1\cdots
i_{p-2k-1}}^{(1)}\right) ,\;\left( V^{(2)i_1\cdots i_{p-2k-1}},\Pi
_{i_1\cdots i_{p-2k-1}}^{(2)}\right) ,\;k=0,\cdots ,a, 
\end{equation}
which we demand to be constrained by 
\begin{equation}
\label{t33}\tilde \gamma _{i_1\ldots i_{p-2k-2}}^{\prime (1)}\equiv \pi
_{i_1\ldots i_{p-2k-2}}^{(1)}\approx 0,\;k=0,\ldots ,c, 
\end{equation}
\begin{equation}
\label{t34}\bar \gamma _{i_1\ldots i_{p-2k-1}}^{\prime (1)}\equiv \Pi
_{i_1\ldots i_{p-2k-1}}^{(1)}\approx 0,\;k=0,\ldots ,a, 
\end{equation}
\begin{equation}
\label{t35}\tilde \gamma _{i_1\ldots i_{p-2k-2}}^{(2)}\equiv -\left(
p-2k-1\right) \pi _{i_1\ldots i_{p-2k-2}}^{(2)}\approx 0,\;k=0,\ldots ,c, 
\end{equation}
\begin{equation}
\label{t36}\bar \gamma _{i_1\ldots i_{p-2k-1}}^{(2)}\equiv -\left(
p-2k\right) \Pi _{i_1\ldots i_{p-2k-1}}^{(2)}\approx 0,\;k=0,\ldots ,a. 
\end{equation}
In the meantime, it is well-known that one can always add to a set of
first-class constraints any combination of first-class constraints whose
coefficients determine an invertible matrix without afflicting the theory.
We notice that from the concrete form of the constraint functions in (\ref
{t26}--\ref{t28}) one can express the momenta $\left( \pi _{i_1\ldots
i_{p-2k-2}}\right) _{k=0,\ldots ,c}$, respectively, $\left( \Pi _{i_1\ldots
i_{p-2k-1}}\right) _{k=0,\ldots ,a}$ under the form 
\begin{equation}
\label{t33a}\pi _{i_1\ldots i_{p-2k-2}}=-\frac 1{\triangle }\left( \partial
^i\tilde \gamma _{ii_1\cdots i_{p-2k-2}}^{(2)}+\frac 1{p-2k-2}\partial
_{\left[ i_1\right. }^{}\tilde \gamma _{\left. i_2\cdots i_{p-2k-2}\right]
}^{(2)}\right) , 
\end{equation}
\begin{equation}
\label{t34a}\Pi _{i_1\ldots i_{p-2k-1}}=-\frac 1{\triangle }\left( \partial
^i\bar \gamma _{ii_1\cdots i_{p-2k-1}}^{(2)}+\frac 1{p-2k-1}\partial
_{\left[ i_1\right. }^{}\bar \gamma _{\left. i_2\cdots i_{p-2k-1}\right]
}^{(2)}\right) . 
\end{equation}
Thus, in view of the above observation, we can redefine the constraints (\ref
{t33}--\ref{t34}) through 
\begin{equation}
\label{t33b}\tilde \gamma _{i_1\ldots i_{p-2k-2}}^{(1)}\equiv \pi
_{i_1\ldots i_{p-2k-2}}-\pi _{i_1\ldots i_{p-2k-2}}^{(1)}\approx
0,\;k=0,\ldots ,c, 
\end{equation}
\begin{equation}
\label{t34b}\bar \gamma _{i_1\ldots i_{p-2k-1}}^{(1)}\equiv \Pi _{i_1\ldots
i_{p-2k-1}}-\Pi _{i_1\ldots i_{p-2k-1}}^{(1)}\approx 0,\;k=0,\ldots ,a. 
\end{equation}
The introduction of the pairs (\ref{t31}--\ref{t32}) is motivated by the
fact that our irreducible Hamiltonian formalism is intended to lead to some
corresponding Lagrangian gauge transformations that are manifestly Lorentz
covariant. Thus, we need to replace the gauge parameters associated with the
first-stage reducibility functions in the reducible context by some other
parameters that render the Lorentz covariance of the Lagrangian gauge
variations of the fields from the irreducible framework. These parameters
are offered precisely by the presence of the supplementary first-class
constraints (\ref{t35}--\ref{t36}) and (\ref{t33b}--\ref{t34b}). As a
consequence of the above redefinitions, the theory having the constraints (%
\ref{t26}--\ref{t28}), (\ref{t25}--\ref{t27}), (\ref{t35}--\ref{t36}) and (%
\ref{t33b}--\ref{t34b}) is still irreducible, first-class, abelian, and has
the same number of physical degrees of freedom like the original model. The
first-class Hamiltonian with respect to these irreducible constraints can be
taken under the form 
\begin{eqnarray}\label{t37}
& &\tilde H^{\prime \prime }=\int d^{2p+1}x\left(
\sum\limits_{k=0}^cA^{i_1\ldots i_{p-2k-2}}\pi _{i_1\ldots
i_{p-2k-2}}^{(2)}+\sum\limits_{k=0}^aH^{i_1\ldots i_{p-2k-1}}\Pi _{i_1\ldots
i_{p-2k-1}}^{(2)}+\right. \nonumber \\  
& &\sum\limits_{k=0}^cB^{(2)i_1\ldots i_{p-2k-2}}\left( \partial ^i\tilde
\gamma _{ii_1\ldots i_{p-2k-2}}^{(2)}+\frac 1{p-2k-2}\partial _{\left[
i_1\right. }\tilde \gamma _{\left. i_2\ldots i_{p-2k-2}\right]
}^{(2)}\right) +\nonumber \\ 
& &\left. \sum\limits_{k=0}^aV^{(2)i_1\ldots i_{p-2k-1}}
\left( \partial ^i\bar
\gamma _{ii_1\ldots i_{p-2k-1}}^{(2)}+\frac 1{p-2k-1}\partial _{\left[
i_1\right. }\bar \gamma _{\left. i_2\ldots i_{p-2k-1}\right] }^{(2)}\right)
\right) +\nonumber \\
& &\tilde H^{\prime }\equiv \int d^{2p+1}x\tilde h^{\prime \prime }.
\end{eqnarray}
In this manner, we constructed an irreducible model (described by the
first-class constraints (\ref{t26}--\ref{t28}), (\ref{t25}--\ref{t27}), (\ref
{t35}--\ref{t36}), (\ref{t33b}--\ref{t34b}) and by the first-class
Hamiltonian (\ref{t37})) associated with topologically coupled $p$- and $%
\left( p+1\right) $-forms.

\subsection{Irreducible Hamiltonian BRST symmetry}

In this subsection we focus on the construction of the Hamiltonian BRST
symmetry for the irreducible model derived in the above. The irreducible
BRST differential $s_I$ has a simple structure due to the abelian character
of the irreducible first-class constraint set, containing only the
irreducible Koszul-Tate operator $\delta $ and the exterior derivative along
the gauge orbits $D$. The irreducible Koszul-Tate complex contains the
fermionic antighost number one minimal antighosts 
\begin{equation}
\label{t38}\left( {\cal P}_{1i_1\cdots i_{p-2k-1}},{\cal P}_{2i_1\cdots
i_{p-2k-1}}\right) _{k=0,\ldots ,a},\left( {\cal P}_{1i_1\cdots i_{p-2k-2}},%
{\cal P}_{2i_1\cdots i_{p-2k-2}}\right) _{k=0,\ldots ,c}, 
\end{equation}
\begin{equation}
\label{t39}\left( P_{1i_1\ldots i_{p-2k}},P_{2i_1\ldots i_{p-2k}}\right)
_{k=0,\ldots ,b},\left( P_{1i_1\ldots i_{p-2k-1}},P_{2i_1\ldots
i_{p-2k-1}}\right) _{k=0,\ldots ,a}, 
\end{equation}
respectively associated with the first-class constraints (\ref{t25}), (\ref
{t26}), (\ref{t33b}), (\ref{t35}), (\ref{t27}), (\ref{t28}), (\ref{t34b})
and (\ref{t36}). The definitions of $\delta $ acting on the variables in the
minimal Koszul-Tate complex take the usual form 
\begin{equation}
\label{br1}\delta z^A=0, 
\end{equation}
\begin{equation}
\label{br2}\delta {\cal P}_{\Delta i_1\cdots i_{p-2k-1}}=-\tilde \gamma
_{i_1\cdots i_{p-2k-1}}^{(\Delta )},\;\Delta =1,2,\;k=0,\ldots ,a, 
\end{equation}
\begin{equation}
\label{br3}\delta {\cal P}_{\Delta i_1\cdots i_{p-2k-2}}=-\tilde \gamma
_{i_1\cdots i_{p-2k-2}}^{(\Delta )},\;\Delta =1,2,\;k=0,\ldots ,c, 
\end{equation}
\begin{equation}
\label{br4}\delta P_{\Delta i_1\ldots i_{p-2k}}=-\bar \gamma _{i_1\ldots
i_{p-2k}}^{(\Delta )},\;\Delta =1,2,\;k=0,\ldots ,b, 
\end{equation}
\begin{equation}
\label{br5}\delta P_{\Delta i_1\ldots i_{p-2k-1}}=-\bar \gamma _{i_1\ldots
i_{p-2k-1}}^{(\Delta )},\;\Delta =1,2,\;k=0,\ldots ,a, 
\end{equation}
where $z^A$ can be any of the original field/momenta or newly added
canonical variable from the pairs (\ref{t13}), (\ref{t14}) or (\ref{t31}--%
\ref{t32}). These definitions ensure the acyclicity at non-vanishing
antighost numbers, as well as the nilpotency of $\delta $, as required by
the BRST formalism. The longitudinal complex involves the minimal ghost
spectrum 
\begin{equation}
\label{t38a}\left( \eta _1^{i_1\cdots i_{p-2k-1}},\eta _2^{i_1\cdots
i_{p-2k-1}}\right) _{k=0,\ldots ,a},\left( \eta _1^{i_1\cdots
i_{p-2k-2}},\eta _2^{i_1\cdots i_{p-2k-2}}\right) _{k=0,\ldots ,c}, 
\end{equation}
\begin{equation}
\label{t39a}\left( C_1^{i_1\ldots i_{p-2k}},C_2^{i_1\ldots i_{p-2k}}\right)
_{k=0,\ldots ,b},\left( C_1^{i_1\ldots i_{p-2k-1}},C_2^{i_1\ldots
i_{p-2k-1}}\right) _{k=0,\ldots ,a}, 
\end{equation}
where all the fields are fermionic, with pure ghost number one, and
respectively correspond to the first-class constraints (\ref{t25}), (\ref
{t26}), (\ref{t33b}), (\ref{t35}), (\ref{t27}), (\ref{t28}), (\ref{t34b}), (%
\ref{t36}). The definitions of $D$ acting on the variables from the
longitudinal complex read as%
\begin{eqnarray}\label{br6}
& &DF=\sum_{\Delta =1}^2\left( \sum\limits_{k=0}^a\left[ F,\tilde \gamma
_{i_1\cdots i_{p-2k-1}}^{(\Delta )}\right] \eta _\Delta ^{i_1\cdots
i_{p-2k-1}}+\right. \nonumber \\
& &\sum\limits_{k=0}^c\left[ F,\tilde \gamma _{i_1\cdots i_{p-2k-2}}^{(\Delta
)}\right] \eta _\Delta ^{i_1\cdots i_{p-2k-2}}+\nonumber \\
& &\left. \sum\limits_{k=0}^b\left[ F,\bar \gamma _{i_1\ldots
i_{p-2k}}^{(\Delta )}\right] C_\Delta ^{i_1\ldots
i_{p-2k}}+\sum\limits_{k=0}^a\left[ F,\bar \gamma _{i_1\ldots
i_{p-2k-1}}^{(\Delta )}\right] C_\Delta ^{i_1\ldots i_{p-2k-1}}\right) ,
\end{eqnarray}
\begin{equation}
\label{br7}D{\cal G}^\Gamma =0, 
\end{equation}
where $F$ is any function of $z^A$, and ${\cal G}^\Gamma $ generically
denotes the minimal ghost spectrum (\ref{t38a}--\ref{t39a}). The exterior
derivative along the gauge orbits is found strongly nilpotent. By enhancing
the action of $\delta $ to the ghosts through 
\begin{equation}
\label{br9}\delta {\cal G}^\Gamma =0, 
\end{equation}
and the action of $D$ to the antighosts (\ref{t38}--\ref{t39}) (which we
globally denote by ${\cal P}_\Gamma $) like 
\begin{equation}
\label{br10}D{\cal P}_\Gamma =0, 
\end{equation}
the homological perturbation theory \cite{hom1}--\cite{hom4} guarantees the
existence of the irreducible Hamiltonian BRST symmetry, $s_I=\delta +D$,
that is nilpotent, $s_I^2=0$. The BRST differential is graded accordingly
the ghost number ($gh$), defined like the difference between the pure ghost
number and the antighost number. This completes the construction of an
irreducible BRST symmetry for the irreducible model deriving from
topologically coupled $p$- and $\left( p+1\right) $-forms. The next step is
to establish the relationship between the irreducible BRST symmetry built
here and the standard reducible Hamiltonian BRST symmetry of the original
model.

\subsection{Classical relationship between the reducible and irreducible
models}

In order to clarify the link between the reducible and irreducible BRST
symmetries we show that the two models are physically equivalent. A simple
count indicates that the numbers of physical (independent) degrees of
freedom of the reducible, respectively, irreducible models coincide. Thus,
we have to investigate only the equality between the sets of physical
observables corresponding to the irreducible and reducible systems. (We
recall that a classical observable is a gauge invariant function.) First, we
show that any observable corresponding to the irreducible model is also an
observable for the reducible one. To this end, we start with an observable $%
F $ of the irreducible theory, that should verify the equations 
\begin{equation}
\label{t28o}\left[ F,\tilde \gamma _{i_1\ldots i_{p-2k-1}}^{(1)}\right]
\approx 0,\;k=0,\cdots ,a,\;\left[ F,\bar \gamma _{i_1\ldots
i_{p-2k}}^{(1)}\right] \approx 0,\;k=0,\cdots ,b, 
\end{equation}
\begin{equation}
\label{t28p}\left[ F,\tilde \gamma _{i_1\ldots i_{p-2k-1}}^{(2)}\right]
\approx 0,\;k=0,\cdots ,a, 
\end{equation}
\begin{equation}
\label{t28q}\left[ F,\bar \gamma _{i_1\ldots i_{p-2k}}^{(2)}\right] \approx
0,\;k=0,\cdots ,b, 
\end{equation}
\begin{equation}
\label{t34bc}\left[ F,\tilde \gamma _{i_1\ldots i_{p-2k-2}}^{(1)}\right]
\approx 0,\;k=0,\cdots ,c,\;\left[ F,\bar \gamma _{i_1\ldots
i_{p-2k-1}}^{(1)}\right] \approx 0,\;k=0,\cdots ,a, 
\end{equation}
\begin{equation}
\label{t34bd}\left[ F,\tilde \gamma _{i_1\ldots i_{p-2k-2}}^{(2)}\right]
\approx 0,\;k=0,\cdots ,c,\;\left[ F,\bar \gamma _{i_1\ldots
i_{p-2k-1}}^{(2)}\right] \approx 0,\;k=0,\cdots ,a. 
\end{equation}
The equations (\ref{t28o}) induce that $F$ does not involve, at least
weakly, the fields $\left( A^{0i_1\ldots i_{p-2k-1}}\right) _{k=0,\cdots ,a}$
and $\left( H^{0i_1\ldots i_{p-2k}}\right) _{k=0,\cdots ,b}$. On the other
hand, the equations (\ref{t34bc}) coupled with the relations (\ref{t33a}--%
\ref{t34a}) and (\ref{t28p}--\ref{t28q}) lead to 
\begin{equation}
\label{t34be}\left[ F,\pi _{i_1\ldots i_{p-2k-2}}^{(1)}\right] \approx
0,\;k=0,\cdots ,c,\;\left[ F,\Pi _{i_1\ldots i_{p-2k-1}}^{(1)}\right]
\approx 0,\;k=0,\cdots ,a. 
\end{equation}
Thus, the equations (\ref{t34bd}) and (\ref{t34be}) indicate that $F$ does
not depend, also at least weakly, on the newly added fields $\left(
B^{(1)i_1\cdots i_{p-2k-2}},B^{(2)i_1\cdots i_{p-2k-2}}\right) _{k=0,\cdots
,c}$ and $\left( V^{(1)i_1\cdots i_{p-2k-1}},V^{(2)i_1\cdots
i_{p-2k-1}}\right) _{k=0,\cdots ,a}$. Let us investigate now the conditions (%
\ref{t28p}) and (\ref{t28q}). For definiteness, we approach here the case $p$
even, the other one being solved in a similar manner. We begin with the last
relation (\ref{t28p}) (assuming that $p$ is even) 
\begin{equation}
\label{t28r}-2\partial _y^i\left[ F\left( x\right) ,\pi _{ii_1}\left(
y\right) \right] -\partial _{i_1}^y\left[ F\left( x\right) ,\pi \left(
y\right) \right] \approx 0. 
\end{equation}
Applying $\partial _y^{i_1}$ on (\ref{t28r}), we infer $-\partial
_y^{i_1}\partial _{i_1}^y\left[ F\left( x\right) ,\pi \left( y\right)
\right] \approx 0$, which further yields 
\begin{equation}
\label{t28s}\left[ F\left( x\right) ,\pi \left( y\right) \right] \approx 0. 
\end{equation}
Substituting (\ref{t28s}) in (\ref{t28r}), we get 
\begin{equation}
\label{t28t}\partial _y^i\left[ F\left( x\right) ,\pi _{ii_1}\left( y\right)
\right] \approx 0. 
\end{equation}
Applying $\partial _y^{i_1}$ on the next relation (\ref{t28p}), namely,%
\begin{eqnarray}\label{t28u}
& &-2\partial _y^i\left[ F\left( x\right) ,\pi _{ii_1i_2i_3}\left( y\right)
\right] -\partial _{i_1}^y\left[ F\left( x\right) ,\pi _{i_2i_3}\left(
y\right) \right] -\partial _{i_2}^y\left[ F\left( x\right) ,\pi
_{i_3i_1}\left( y\right) \right] -\nonumber \\ 
& &-\partial _{i_3}^y\left[ F\left( x\right) ,\pi _{i_1i_2}\left(
y\right) \right] \approx 0,
\end{eqnarray}
and using (\ref{t28t}), we derive $-\partial _y^{i_1}\partial _{i_1}^y\left[
F\left( x\right) ,\pi _{i_2i_3}\left( y\right) \right] \approx 0$, hence 
\begin{equation}
\label{t28v}\left[ F\left( x\right) ,\pi _{i_2i_3}\left( y\right) \right]
\approx 0. 
\end{equation}
Replacing the above result in (\ref{t28u}) and reprising the same program on
the next relations (\ref{t28p}), we are led to 
\begin{equation}
\label{t28x}\left[ F\left( x\right) ,\pi _{i_1\cdots i_{p-2k-2}}\left(
y\right) \right] \approx 0,\;k=0,\cdots ,c, 
\end{equation}
which, inserted into the first equation (\ref{t28p}), yield 
\begin{equation}
\label{t28y}\left[ F\left( x\right) ,\tilde G_{i_1\cdots
i_{p-1}}^{(2)}\left( y\right) \right] \approx 0. 
\end{equation}
If we act along the same line, but starting from the last equation (\ref
{t28q}), we will accordingly arrive at 
\begin{equation}
\label{t28w}\left[ F\left( x\right) ,\Pi _{i_1\cdots i_{p-2k-1}}\left(
y\right) \right] \approx 0,\;k=0,\cdots ,a, 
\end{equation}
which, substituted in (\ref{t28q}) for $k=0$ imply 
\begin{equation}
\label{t28z}\left[ F\left( x\right) ,\bar G_{i_1\cdots i_p}^{(2)}\left(
y\right) \right] \approx 0. 
\end{equation}
The equations (\ref{t28x}) and (\ref{t28w}) indicate that $F$ does not
depend, at least weakly, on the fields $\left( A^{i_1\ldots
i_{p-2k-2}}\right) _{k=0,\cdots ,c}$ and $\left( H^{i_1\ldots
i_{p-2k-1}}\right) _{k=0,\cdots ,a}$. We can summarize the prior results by
stating that if $F$ denotes an observable of the irreducible theory, then it
does not depend on any of the newly introduced fields (\ref{t13}), (\ref{t14}%
) and (\ref{t31}--\ref{t32}). Moreover, it satisfies the relations 
\begin{equation}
\label{t26n}\left[ F\left( x\right) ,\tilde G_{i_1\cdots
i_{p-1}}^{(1)}\left( y\right) \right] \approx 0,\;\left[ F\left( x\right)
,\bar G_{i_1\cdots i_p}^{(1)}\left( y\right) \right] \approx 0, 
\end{equation}
(see (\ref{t28o}) for $k=0$), and also (\ref{t28y}), (\ref{t28z}), which are
precisely the equations verified by an observable of the reducible model.
All these show that if $F$ is an observable of the irreducible theory, then
it is also an observable of the redundant system. The converse is valid,
too, because any observable of the redundant model checks the equations (\ref
{t28y}), (\ref{t28z}--\ref{t26n}), and does not depend on the newly added
canonical pairs, such that (\ref{t28o}--\ref{t34bd}) are automatically
satisfied. Thus, as both the irreducible and reducible models display the
same physical observables, the zeroth order cohomological groups of the
reducible and irreducible BRST symmetries, $s_R$ and $s_I$, are equal 
\begin{equation}
\label{eqb}H^0\left( s_R\right) =H^0\left( s_I\right) . 
\end{equation}
In view of this, the reducible and irreducible models are equivalent from
the BRST formalism point of view, i.e., from the point of view of the basic
requirements of the BRST symmetry, $s^2=0$ and $H^0\left( s\right) =\left\{ 
{\rm physical\;observables}\right\} $. As a consequence, we can substitute
the reducible Hamiltonian BRST symmetry for the original system by that of
the irreducible theory. This further implies that at the BRST quantization
level we can also replace the Hamiltonian BRST quantization of topologically
coupled abelian $p$- and $\left( p+1\right) $-forms with that of the
irreducible first-class theory.

\subsection{Hamiltonian BRST quantization of the irreducible theory}

In the sequel we rely on the last conclusion and investigate the Hamiltonian
BRST quantization of the irreducible model. The minimal antighost and ghost
spectra are offered by (\ref{t38}--\ref{t39}) and (\ref{t38a}--\ref{t39a}).
It is convenient to work with the non-minimal sector 
\begin{equation}
\label{t42}\left( P_{\bar \eta }^{i_1\ldots i_{p-2k-1}},\bar \eta
_{i_1\ldots i_{p-2k-1}}\right) ,\;\left( P_{\bar \eta ^1}^{i_1\ldots
i_{p-2k-1}},\bar \eta _{i_1\ldots i_{p-2k-1}}^1\right) ,\;k=0,\ldots ,a, 
\end{equation}
\begin{equation}
\label{t43}\left( P_b^{i_1\ldots i_{p-2k-1}},b_{i_1\ldots i_{p-2k-1}}\right)
,\;\left( P_{b^1}^{i_1\ldots i_{p-2k-1}},b_{i_1\ldots i_{p-2k-1}}^1\right)
,\;k=0,\ldots ,a, 
\end{equation}
\begin{equation}
\label{t44}\left( P_{\bar C}^{i_1\ldots i_{p-2k}},\bar C_{i_1\ldots
i_{p-2k}}\right) ,\;\left( P_{\bar C^1}^{i_1\ldots i_{p-2k}},\bar
C_{i_1\ldots i_{p-2k}}^1\right) ,\;k=0,\ldots ,b, 
\end{equation}
\begin{equation}
\label{t45}\left( P_{\tilde b}^{i_1\ldots i_{p-2k}},\tilde b_{i_1\ldots
i_{p-2k}}\right) ,\;\left( P_{\tilde b^1}^{i_1\ldots i_{p-2k}},\tilde
b_{i_1\ldots i_{p-2k}}^1\right) ,\;k=0,\ldots ,b. 
\end{equation}
The variables (\ref{t43}), (\ref{t45}) are bosonic and have ghost number
zero. The fields from (\ref{t42}), (\ref{t44}) are fermionic, the $P$'s
possessing ghost number one, while the $\bar \eta $'s and $\bar C$'s have
ghost number minus one. The non-minimal BRST canonical generator of the
irreducible Hamiltonian BRST symmetry reads as%
\begin{eqnarray}\label{t46}
& &\tilde \Omega =\int d^{2p+1}x\left( \sum_{\Delta =1}^2\left(
\sum_{k=1}^p\eta _\Delta ^{i_1\ldots i_{p-k}}\tilde \gamma _{i_1\ldots
i_{p-k}}^{(\Delta )}+\sum_{k=1}^{p+1}C_\Delta ^{i_1\ldots i_{p-k+1}}\bar
\gamma _{i_1\ldots i_{p-k+1}}^{(\Delta )}\right) +\right. \nonumber \\ 
& &\sum_{k=0}^a\left( P_{\bar \eta }^{i_1\ldots i_{p-2k-1}}b_{i_1\ldots
i_{p-2k-1}}+P_{\bar \eta ^1}^{i_1\ldots i_{p-2k-1}}b_{i_1\ldots
i_{p-2k-1}}^1\right) +\nonumber \\ 
& &\left. \sum_{k=0}^b\left( P_{\bar C}^{i_1\ldots i_{p-2k}}\tilde
b_{i_1\ldots i_{p-2k}}+P_{\bar C^1}^{i_1\ldots i_{p-2k}}\tilde b_{i_1\ldots
i_{p-2k}}^1\right) \right) , 
\end{eqnarray}
while the BRST-invariant extension of $\tilde H^{\prime \prime }$ has the
form 
\begin{eqnarray}\label{t47}
& &\tilde H_B^{\prime \prime }=\tilde H^{\prime \prime }+
\int d^{2p+1}x\left(
\sum_{k=0}^a\eta _1^{i_1\ldots i_{p-2k-1}}{\cal P}_{2i_1\ldots
i_{p-2k-1}}+\frac 1pC_2^{i_1\ldots i_p}\partial _{\left[ i_1\right.
}P_{2\left. i_2\ldots i_p\right] }-\right. \nonumber \\
& &\sum_{k=0}^c\frac 1{p-2k-1}\eta _1^{i_1\ldots i_{p-2k-2}}{\cal P}%
_{2i_1\ldots i_{p-2k-2}}+\frac 1{p-1}\eta _2^{i_1\ldots i_{p-1}}\partial
_{\left[ i_1\right. }{\cal P}_{2\left. i_2\ldots i_{p-1}\right] }+
\nonumber \\ 
& &\sum_{k=1}^a\eta _2^{i_1\ldots i_{p-2k-1}}\left(
\frac{p-2k}{p-2k+1}\partial
^i{\cal P}_{2ii_1\ldots i_{p-2k-1}}+\frac 1{p-2k-1}\partial _{\left[
i_1\right. }{\cal P}_{2\left. i_2\ldots i_{p-2k-1}\right] }\right) - 
\nonumber \\
& &\sum_{k=0}^c\eta _2^{i_1\ldots i_{p-2k-2}}\left( \left( p-2k-1\right)
\partial ^i{\cal P}_{2ii_1\ldots i_{p-2k-2}}+\frac{p-2k-1}{p-2k-2}\partial
_{\left[ i_1\right. }{\cal P}_{2\left. i_2\ldots i_{p-2k-2}\right] }\right)
+\nonumber \\
& &\sum_{k=0}^bC_1^{i_1\ldots i_{p-2k}}P_{2i_1\ldots
i_{p-2k}}-\sum_{k=0}^a\frac 1{p-2k}C_1^{i_1\ldots i_{p-2k-1}}P_{2i_1\ldots
i_{p-2k-1}}+\nonumber \\
& &\sum_{k=1}^bC_2^{i_1\ldots i_{p-2k}}\left( \frac{p-2k+1}{p-2k+2}\partial
^iP_{2ii_1\ldots i_{p-2k}}+\frac 1{p-2k}\partial _{\left[ i_1\right.
}P_{2\left. i_2\ldots i_{p-2k}\right] }\right) -\nonumber \\
& &\left. \sum_{k=0}^aC_2^{i_1\ldots i_{p-2k-1}}\left( \left(
p-2k\right) \partial ^iP_{2ii_1\ldots i_{p-2k-1}}+\frac{p-2k}{p-2k-1}%
\partial _{\left[ i_1\right. }P_{2\left. i_2\ldots i_{p-2k-1}\right]
}\right) \right) . 
\end{eqnarray}
In order to fix the gauge we choose the gauge-fixing fermion 
\begin{eqnarray}\label{t48}
& &\tilde K=\int d^{2p+1}x\left( {\cal P}_{1i_1\ldots i_{p-1}}\left( \partial
_iA^{ii_1\ldots i_{p-1}}+\frac 1{p-1}\partial ^{\left[ i_1\right.
}B^{(1)}{}^{\left. i_2\ldots i_{p-1}\right] }\right) +\right. \nonumber \\
& &\sum_{k=1}^a{\cal P}_{1i_1\ldots i_{p-2k-1}}\left( \partial
_iB^{(1)ii_1\ldots i_{p-2k-1}}+\frac 1{p-2k-1}\partial ^{\left[ i_1\right.
}B^{(1)\left. i_2\ldots i_{p-2k-1}\right] }\right) +\nonumber \\
& &\left( -\right) ^{p+1}\sum_{k=0}^c{\cal P}_{1i_1\ldots i_{p-2k-2}}\left(
\left( p-2k-1\right) \partial _iA^{ii_1\ldots i_{p-2k-2}0}+\partial ^{\left[
i_1\right. }A^{\left. i_2\ldots i_{p-2k-2}\right] 0}\right) +\nonumber \\
& &\sum_{k=0}^aP_{b^1}^{i_1\ldots i_{p-2k-1}}\left( {\cal P}_{1i_1\ldots
i_{p-2k-1}}-\bar \eta _{i_1\ldots i_{p-2k-1}}+\stackrel{.}{\bar \eta }%
_{i_1\ldots i_{p-2k-1}}^1\right) +\nonumber \\
& &\sum_{k=0}^aP_b^{i_1\ldots i_{p-2k-1}}\left( \bar \eta _{i_1\ldots
i_{p-2k-1}}^1+\stackrel{.}{\bar \eta }_{i_1\ldots i_{p-2k-1}}\right) +
\nonumber \\
& &P_{1i_1\ldots i_p}\left( \partial _iH^{ii_1\ldots i_p}+\frac 1p\partial
^{\left[ i_1\right. }V^{(1)}{}^{\left. i_2\ldots i_p\right] }\right) +
\nonumber \\
& &\sum_{k=1}^bP_{1i_1\ldots i_{p-2k}}\left( \partial _iV^{(1)}{}^{ii_1\ldots
i_{p-2k}}+\frac 1{p-2k}\partial ^{\left[ i_1\right. }V^{(1)\left. i_2\ldots
i_{p-2k}\right] }\right) +\nonumber \\
& &\left( -\right) ^p\sum_{k=0}^aP_{1i_1\ldots i_{p-2k-1}}\left( \left(
p-2k\right) \partial _iH^{ii_1\ldots i_{p-2k-1}0}+\partial ^{\left[
i_1\right. }H^{\left. i_2\ldots i_{p-2k-1}\right] 0}\right) +\nonumber \\
& &\sum_{k=0}^bP_{\tilde b^1}^{i_1\ldots i_{p-2k}}\left( P_{1i_1\ldots
i_{p-2k}}-\bar C_{i_1\ldots i_{p-2k}}+\stackrel{.}{\bar C}_{i_1\ldots
i_{p-2k}}^1\right) +\nonumber \\ 
& &\left. \sum_{k=0}^bP_{\tilde b}^{i_1\ldots i_{p-2k}}\left( \bar
C_{i_1\ldots i_{p-2k}}^1+\stackrel{.}{\bar C}_{i_1\ldots i_{p-2k}}\right)
\right) . 
\end{eqnarray}
The corresponding path integral, resulting after some computation, will be%
\begin{eqnarray}\label{t49}
& &Z_{\tilde K}=
\int {\cal D}H^{\mu _1\ldots \mu _{p+1}}{\cal D}A^{\mu _1\ldots
\mu _p}\times \nonumber \\
& &\left( \prod_{k=0}^a{\cal D}V^{(1)\mu _1
\ldots \mu _{p-2k-1}}\right) \left(
\prod_{k=0}^c{\cal D}B^{(1)\mu _1\ldots \mu _{p-2k-2}}\right) \times  
\nonumber \\
& &\left( \prod_{k=0}^b\left( {\cal D}\tilde
b_{\mu _1\ldots \mu _{p-2k}}{\cal D%
}C_2^{\mu _1\ldots \mu _{p-2k}}{\cal D}\bar C_{\mu _1\ldots \mu
_{p-2k}}\right) \right) \times \nonumber \\
& &\left( \prod_{k=0}^a\left( {\cal D}b_{\mu _1\ldots \mu _{p-2k-1}}%
{\cal D}\eta _2^{\mu _1\ldots \mu _{p-2k-1}}{\cal D}\bar \eta _{\mu _1\ldots
\mu _{p-2k-1}}\right) \right) \exp iS_{\tilde K},
\end{eqnarray}
where 
\begin{eqnarray}\label{t50}
& &S_{\tilde K}=\tilde S_0^L+
\int d^{2p+2}x\left( -\sum_{k=0}^a\bar \eta _{\mu
_1\ldots \mu _{p-2k-1}}\Box \eta _2^{\mu _1\ldots \mu
_{p-2k-1}}-\right. \nonumber \\
& &\sum_{k=0}^b\bar C_{\mu _1\ldots \mu _{p-2k}}\Box C_2^{\mu
_1\ldots \mu _{p-2k}}+ \nonumber \\
& &b_{\mu _1\ldots \mu _{p-1}}\left( \partial _\mu A^{\mu \mu _1\ldots \mu
_{p-1}}+\frac 1{p-1}\partial ^{\left[ \mu _1\right. }B^{(1)\left. \mu
_2\ldots \mu _{p-1}\right] }\right) +\nonumber \\
& &\sum_{k=1}^ab_{\mu _1\ldots \mu _{p-2k-1}}
\left( \partial _\mu B^{(1)\mu \mu
_1\ldots \mu _{p-2k-1}}+\frac 1{p-2k-1}\partial ^{\left[ \mu _1\right.
}B^{(1)\left. \mu _2\ldots \mu _{p-2k-1}\right] }\right) +\nonumber \\
& &\tilde b_{\mu _1\ldots \mu _p}\left( \partial _\mu H^{\mu \mu _1\ldots \mu
_p}+\frac 1p\partial ^{\left[ \mu _1\right. }V^{(1)\left. \mu _2\ldots \mu
_p\right] }\right) +\nonumber \\
& &\left. \sum_{k=1}^b\tilde b_{\mu _1\ldots \mu _{p-2k}}\left(
\partial _\mu V^{(1)\mu \mu _1\ldots \mu _{p-2k}}+\frac 1{p-2k}\partial
^{\left[ \mu _1\right. }V^{(1)\left. \mu _2\ldots \mu _{p-2k}\right]
}\right) \right) , 
\end{eqnarray}
and $\Box =\partial ^\mu \partial _\mu $. The action $\tilde S_0^L$ is
nothing but the original action, expressed by (\ref{t1}). We mention that in
obtaining (\ref{t49}--\ref{t50}) we performed the identifications 
\begin{equation}
\label{t52}B^{(1)\mu _1\ldots \mu _{p-2k-2}}\equiv \left( A^{0i_1\cdots
i_{p-2k-3}},B^{(1)i_1\cdots i_{p-2k-2}}\right) ,\;k=0,\cdots ,c, 
\end{equation}
\begin{equation}
\label{t53}b_{\mu _1\ldots \mu _{p-2k-1}}\equiv \left( \pi _{i_1\ldots
i_{p-2k-2}}^{(1)},b_{i_1\ldots i_{p-2k-1}}\right) ,\;k=0,\ldots ,a, 
\end{equation}
\begin{equation}
\label{t54}\eta _2^{\mu _1\ldots \mu _{p-2k-1}}\equiv \left( \eta
_2^{i_1\ldots i_{p-2k-2}},\eta _2^{i_1\ldots i_{p-2k-1}}\right)
,\;k=0,\ldots ,a, 
\end{equation}
\begin{equation}
\label{t55}\bar \eta _{\mu _1\ldots \mu _{p-2k-1}}\equiv \left( -{\cal P}%
_{1i_1\ldots i_{p-2k-2}},\bar \eta _{i_1\ldots i_{p-2k-1}}\right)
,\;k=0,\ldots ,a, 
\end{equation}
\begin{equation}
\label{t56}V^{(1)\mu _1\ldots \mu _{p-2k-1}}\equiv \left( H^{0i_1\cdots
i_{p-2k-2}},V^{(1)i_1\cdots i_{p-2k-1}}\right) ,\;k=0,\cdots ,a, 
\end{equation}
\begin{equation}
\label{t57}\tilde b_{\mu _1\ldots \mu _{p-2k}}\equiv \left( \Pi _{i_1\ldots
i_{p-2k-1}}^{(1)},\tilde b_{i_1\ldots i_{p-2k}}\right) ,\;k=0,\ldots ,b, 
\end{equation}
\begin{equation}
\label{t58}C_2^{\mu _1\ldots \mu _{p-2k}}\equiv \left( C_2^{i_1\ldots
i_{p-2k-1}},C_2^{i_1\ldots i_{p-2k}}\right) ,\;k=0,\ldots ,b, 
\end{equation}
\begin{equation}
\label{t59}\bar C_{\mu _1\ldots \mu _{p-2k}}\equiv \left( -P_{1i_1\ldots
i_{p-2k-1}},\bar C_{i_1\ldots i_{p-2k}}\right) ,\;k=0,\ldots ,b. 
\end{equation}
It is easy to check that the gauge-fixed action (\ref{t50}) has no residual
gauge invariances. Hence, following our irreducible treatment, we inferred a
path integral for topologically coupled $p$- and $\left( p+1\right) $-form
gauge fields that involves no ghosts for ghosts, and, in addition, is
Lorentz covariant.

\section{Irreducible treatment for interacting theories with topological
coupling}

In the sequel we extend our irreducible treatment to interacting gauge
theories with topological coupling. A possibility would be to investigate
the canonical analysis of the interacting theory and then develop an
irreducible method along the lines exposed in the previous section. A major
difficulty in implementing this program is that the interaction terms may
involve higher order derivatives of the fields, which would make the
canonical approach too complicated. An alternative that surpasses this
inconvenient is to analyze whether our irreducible Hamiltonian procedure
induces a corresponding irreducible Lagrangian version, and, if the answer
is affirmative, to solve the interacting case within the irreducible
Lagrangian context. We will see that this idea can be consistently enforced,
our irreducible Hamiltonian scheme for topologically coupled $p$- and $%
\left( p+1\right) $-form gauge fields allowing indeed an irreducible
Lagrangian formalism that maintains the space-time locality and Lorentz
covariance of the resulting gauge-fixed action. The manifest covariance will
be restored precisely due to the introduction in the theory of the
supplementary canonical pairs (\ref{t31}--\ref{t32}). While in the
Hamiltonian background the distinction between primary and secondary
constraints is not significant, this aspect becomes important at the
Lagrangian level in order to obtain the gauge transformations of the
Lagrangian action. This is why in what follows we work with a model of
irreducible Hamiltonian theory in the case of topological coupling in the
framework of which we assume that (\ref{t25}), (\ref{t27}), (\ref{t33b}--\ref
{t34b}) are primary constraints whose consistencies respectively imply the
secondary ones (\ref{t26}), (\ref{t28}), (\ref{t35}--\ref{t36}). The
derivation of the gauge transformations of our irreducible model involves
three steps. First, we write down the associated extended action 
\begin{eqnarray}\label{ext1}
& &\tilde S_0^{\prime \prime E}=\int d^{2p+2}x\left( \sum\limits_{k=0}^b\dot
A^{j_1\ldots j_{p-2k}}\pi _{j_1\ldots j_{p-2k}}+\sum\limits_{k=0}^{a+1}\dot
H^{i_1\ldots i_{p-2k+1}}\Pi _{i_1\ldots i_{p-2k+1}}+\right. \nonumber \\ 
& &\sum\limits_{k=0}^d\dot A^{0i_1\ldots i_{p-2k-3}}\pi _{0i_1\ldots
i_{p-2k-3}}+\sum\limits_{k=0}^c\dot H^{0i_1\ldots i_{p-2k-2}}\Pi
_{0i_1\ldots i_{p-2k-2}}+\nonumber \\
& &\sum\limits_{\Delta =1}^2\sum\limits_{k=0}^c\dot B^{(\Delta )i_1\cdots
i_{p-2k-2}}\pi _{i_1\cdots i_{p-2k-2}}^{(\Delta )}+\sum\limits_{\Delta
=1}^2\sum\limits_{k=0}^a\dot V^{(\Delta )i_1\cdots i_{p-2k-1}}\Pi _{i_1\cdots
i_{p-2k-1}}^{(\Delta)}-\tilde h^{\prime \prime }-\nonumber \\
& &\sum\limits_{\Delta =1}^2\sum\limits_{k=0}^a\tilde \gamma _{i_1\ldots
i_{p-2k-1}}^{(\Delta )}\tilde u^{(\Delta )i_1\ldots
i_{p-2k-1}}-\sum\limits_{\Delta =1}^2\sum\limits_{k=0}^b\bar \gamma
_{i_1\ldots i_{p-2k}}^{(\Delta )}\bar u^{(\Delta )i_1\ldots i_{p-2k}}- 
\nonumber \\
& &\left. \sum\limits_{\Delta =1}^2\sum\limits_{k=0}^c\tilde \gamma
_{i_1\ldots i_{p-2k-2}}^{(\Delta )}\tilde u^{(\Delta )i_1\ldots
i_{p-2k-2}}-\sum\limits_{\Delta =1}^2\sum\limits_{k=0}^a\bar \gamma
_{i_1\ldots i_{p-2k-1}}^{(\Delta )}\bar u^{(\Delta )i_1\ldots
i_{p-2k-1}}\right) , 
\end{eqnarray}
and determine its gauge invariances. In the last relation $\tilde h^{\prime
\prime }$ is given by (\ref{t37}), while the $\tilde u^{(\Delta )}$'s and $%
\bar u^{(\Delta )}$'s represent the Lagrange multipliers of the
corresponding constraints. Second, on the one hand with the help of the
extended action (\ref{ext1}) we infer the so-called total action by setting
zero all the multipliers carrying the index $(2)$ (and associated by virtue
of our choice with the secondary constraints of the irreducible model), and,
on the other hand, we determine the gauge invariances of the total action by
taking all the gauge variations of the multipliers associated with the
secondary constraints to vanish. Third, we deduce the Lagrangian action for
the irreducible model together with its gauge invariances by eliminating all
the momenta and the remaining Lagrange multipliers on their equations of
motion resulting from the total formalism. In addition, we notice that the
fields carrying the superscript $(2)$ and also $\left( A^{j_1\ldots
j_{p-2k-2}}\right) _{k=0,\cdots ,c}$, $\left( H^{i_1\ldots
i_{p-2k-1}}\right) _{k=0,\cdots ,a}$ are auxiliary variables, hence we can
remove them from the irreducible model. As a result of this three-step
algorithm, we get that the Lagrangian action implied by the irreducible
Hamiltonian theory is nothing but the original action 
\begin{eqnarray}\label{ext2}
& &\tilde S_0^{\prime \prime L}
\left[ A^{\mu _1\ldots \mu _p},H^{\mu _1\ldots
\mu _{p+1}},B^{(1)\mu _1\ldots \mu _{p-2k-2}},V^{(1)\mu _1\ldots \mu
_{p-2k-1}}\right] =\nonumber \\
& &\tilde S_0^L\left[ A^{\mu _1\ldots \mu _p},H^{\mu _1\ldots \mu
_{p+1}}\right] ,
\end{eqnarray}
while the corresponding gauge transformations, which can be checked to be
irreducible, are expressed by 
\begin{equation}
\label{t60}\delta _\epsilon A^{\mu _1\ldots \mu _p}=\partial ^{\left[ \mu
_1\right. }\tilde \epsilon ^{\left. \mu _2\ldots \mu _p\right] },
\end{equation}
\begin{eqnarray}\label{t61}
& &\delta _\epsilon B^{(1)\mu _1\ldots \mu _{p-2k-2}}=\partial ^{\left[ \mu
_1\right. }\tilde \epsilon ^{\left. \mu _2\ldots \mu _{p-2k-2}\right] }+
\nonumber \\
& &\left( p-2k-1\right) \partial _\mu \tilde \epsilon ^{\mu \mu
_1\ldots \mu _{p-2k-2}},\;k=0,\cdots ,c,
\end{eqnarray}
\begin{equation}
\label{t62}\delta _\epsilon H^{\mu _1\ldots \mu _{p+1}}=\partial ^{\left[
\mu _1\right. }\bar \epsilon ^{\left. \mu _2\ldots \mu _{p+1}\right] },
\end{equation}
\begin{eqnarray}\label{t63}
& &\delta _\epsilon V^{(1)\mu _1\ldots \mu _{p-2k-1}}=\partial ^{\left[ \mu
_1\right. }\bar \epsilon ^{\left. \mu _2\ldots \mu _{p-2k-1}\right] }+
\nonumber \\
& &\left( p-2k\right) \partial _\mu \bar \epsilon ^{\mu \mu _1\ldots
\mu _{p-2k-1}},\;k=0,\cdots ,a, 
\end{eqnarray}
where the identifications (\ref{t52}) and (\ref{t56}) have also been
employed. The gauge parameters involved with (\ref{t60}--\ref{t63}) are
defined by 
\begin{equation}
\label{t64}\tilde \epsilon ^{\mu _1\ldots \mu _{p-2k-1}}\equiv \left( \tilde
\epsilon ^{i_1\cdots i_{p-2k-2}},\tilde \epsilon ^{i_1\cdots
i_{p-2k-1}}\right) ,\;k=0,\cdots ,a,
\end{equation}
\begin{equation}
\label{t65}\bar \epsilon ^{\mu _1\ldots \mu _{p-2k}}\equiv \left( \bar
\epsilon ^{i_1\cdots i_{p-2k-1}},\bar \epsilon ^{i_1\cdots i_{p-2k}}\right)
,\;k=0,\cdots ,b,
\end{equation}
where the parameters $\left( \tilde \epsilon ^{i_1\cdots i_{p-2k-2}},\tilde
\epsilon ^{i_1\cdots i_{p-2k-1}}\right) $ correspond to the constraints (\ref
{t35}), respectively, (\ref{t26}), while $\left( \bar \epsilon ^{i_1\cdots
i_{p-2k-1}},\bar \epsilon ^{i_1\cdots i_{p-2k}}\right) $ are associated with
the constraints (\ref{t36}), respectively, (\ref{t28}). We remark that the
gauge variations (\ref{t60}) and (\ref{t62}) involved with the original
fields $A^{\mu _1\ldots \mu _p}$ and $H^{\mu _1\ldots \mu _{p+1}}$ are
nothing but the gauge invariances of the original action (\ref{t1}).
However, although these transformations alone are reducible, the entire set
of gauge transformations (\ref{t60}--\ref{t63}) connected to the larger
field spectrum is irreducible. In this manner we constructed an irreducible
Lagrangian model originating in our irreducible Hamiltonian approach
addressed in the previous section. It can be shown that we can recover the
relations (\ref{t49}--\ref{t50}) in the framework of the antifield-BRST
quantization of this irreducible Lagrangian model by using an appropriate
non-minimal sector and gauge-fixing fermion. The non-minimal solution to the
master equation for the irreducible Lagrangian system reads as%
\begin{eqnarray}\label{ext3}
& &\tilde S^{\prime \prime }=
\tilde S_0^L+\int d^{2p+2}x\left( A_{\mu _1\ldots
\mu _p}^{*}\partial ^{\left[ \mu _1\right. }\eta ^{\left. \mu _2\ldots \mu
_p\right] }+H_{\mu _1\ldots \mu _{p+1}}^{*}\partial ^{\left[ \mu _1\right.
}C^{\left. \mu _2\ldots \mu _{p+1}\right] }+\right. \nonumber \\
& &\sum_{k=0}^cB_{\mu _1\ldots \mu _{p-2k-2}}^{*(1)}\left( \partial ^{\left[
\mu _1\right. }\eta ^{\left. \mu _2\ldots \mu _{p-2k-2}\right] }+\left(
p-2k-1\right) \partial _\mu \eta ^{\mu \mu _1\ldots \mu _{p-2k-2}}\right) + 
\nonumber \\
& &\sum_{k=0}^aV_{\mu _1\ldots \mu _{p-2k-1}}^{*(1)}\left( \partial ^{\left[
\mu _1\right. }C^{\left. \mu _2\ldots \mu _{p-2k-1}\right] }+\left(
p-2k\right) \partial _\mu C^{\mu \mu _1\ldots \mu _{p-2k-1}}\right) - 
\nonumber \\
& &\left. \sum_{k=0}^a\bar \eta _{\mu _1\ldots \mu
_{p-2k-1}}^{*}b^{\mu _1\ldots \mu _{p-2k-1}}-\sum_{k=0}^b\bar C_{\mu
_1\ldots \mu _{p-2k}}^{*}\tilde b^{\mu _1\ldots \mu _{p-2k}}\right) , 
\end{eqnarray}
where $\left( \eta ^{\mu _1\ldots \mu _{p-2k-1}}\right) _{k=0,\cdots ,a}$
and $\left( C^{\mu _1\ldots \mu _{p-2k}}\right) _{k=0,\cdots ,b}$ signify
the Lagrangian pure ghost number one ghosts, the star variables denote the
antifields of the corresponding fields, and the other variables belong to
the non-minimal sector. If we choose a gauge-fixing fermion of the type 
\begin{equation}
\label{ext4}\psi =-\int d^{2p+2}x\left( \sum_{k=0}^a\tilde \chi ^{\mu
_1\ldots \mu _{p-2k-1}}\bar \eta _{\mu _1\ldots \mu
_{p-2k-1}}+\sum_{k=0}^b\bar \chi ^{\mu _1\ldots \mu _{p-2k}}\bar C_{\mu
_1\ldots \mu _{p-2k}}\right) ,
\end{equation}
where the functions $\tilde \chi ^{\mu _1\ldots \mu _{p-2k-1}}$ and $\bar
\chi ^{\mu _1\ldots \mu _{p-2k}}$ are expressed by 
\begin{equation}
\label{ext5}\tilde \chi ^{\mu _1\ldots \mu _{p-1}}=\partial _\mu A^{\mu \mu
_1\ldots \mu _{p-1}}+\frac 1{p-1}\partial ^{\left[ \mu _1\right.
}B^{(1)\left. \mu _2\ldots \mu _{p-1}\right] },
\end{equation}
\begin{equation}
\label{ext6}\tilde \chi ^{\mu _1\ldots \mu _{p-2k-1}}=\partial _\mu
B^{(1)\mu \mu _1\ldots \mu _{p-2k-1}}+\frac 1{p-2k-1}\partial ^{\left[ \mu
_1\right. }B^{(1)\left. \mu _2\ldots \mu _{p-2k-1}\right] },\;k=1,\cdots ,a,
\end{equation}
\begin{equation}
\label{ext7}\bar \chi ^{\mu _1\ldots \mu _p}=\partial _\mu H^{\mu \mu
_1\ldots \mu _p}+\frac 1p\partial ^{\left[ \mu _1\right. }V^{(1)\left. \mu
_2\ldots \mu _p\right] },
\end{equation}
\begin{equation}
\label{ext8}\bar \chi ^{\mu _1\ldots \mu _{p-2k}}=\partial _\mu V^{(1)\mu
\mu _1\ldots \mu _{p-2k}}+\frac 1{p-2k}\partial ^{\left[ \mu _1\right.
}V^{(1)\left. \mu _2\ldots \mu _{p-2k}\right] },\;k=1,\cdots ,b,
\end{equation}
and eliminate all the antifields from (\ref{ext3}) with the help of (\ref
{ext4}) we are led precisely to (\ref{t49}--\ref{t50}) modulo the
identifications 
\begin{equation}
\label{ext9}\eta ^{\mu _1\ldots \mu _{p-2k-1}}\equiv \eta _2^{\mu _1\ldots
\mu _{p-2k-1}},
\end{equation}
\begin{equation}
\label{ext10}C^{\mu _1\ldots \mu _{p-2k}}\equiv C_2^{\mu _1\ldots \mu
_{p-2k}}.
\end{equation}
In consequence, we emphasized how our irreducible Hamiltonian procedure
gives rise to an irreducible covariant Lagrangian approach for topologically
coupled $p$- and $\left( p+1\right) $-form gauge fields that outputs the
path integral derived in the Hamiltonian context. Taking into consideration
this result, the interaction case can be solved in a direct manner. Indeed,
if one adds to the Lagrangian action (\ref{t1}) some interaction terms which
are invariant under the original reducible gauge transformations (\ref{t60})
and (\ref{t62}), then the starting point toward an irreducible Lagrangian
approach to the interacting system is represented by the interacting
Lagrangian action subject to the irreducible gauge transformations (\ref{t60}%
--\ref{t63}) of the broader field spectrum. The main point is that even if
the interaction terms involve higher-order derivatives of the fields, this
does not afflict in any way our procedure as the interacting Lagrangian
action satisfies the same Noether identities like in the absence of the
interaction. Therefore, the non-minimal solution to the master equation
results from (\ref{ext3}) in which we replace $\tilde S_0^L$ with the action
of the interacting Lagrangian model under study. Consequently, we can still
employ the gauge-fixing fermion (\ref{ext4}), which will produce a
gauge-fixed action of the type (\ref{t50}) excepting the starting Lagrangian
action that must contain the gauge-invariant interaction terms. Moreover,
our formalism can yet be extended to interacting theories like the ones
discussed above which contain more sorts of abelian $p$-form gauge fields.
These theories are important in order to derive all consistent interactions
between $p$-form gauge fields \cite{21}. In this light, our irreducible
Hamiltonian procedure gives rise to an irreducible Lagrangian approach which
proves to be efficient at the irreducible investigation of interacting
theories with topological coupling.

\section{Conclusion}

In this paper we develop a consistent irreducible Hamiltonian BRST treatment
of $p$-form gauge theories with topological coupling. We start with a
quadratic action describing topologically coupled abelian $p$- and $\left(
p+1\right) $-form gauge fields and construct an irreducible Hamiltonian
first-class model that is equivalent at the BRST quantization level with the
starting redundant theory. The irreducibility is enforced in the background
of the Koszul-Tate complex via making all the initial antighost number one
co-cycles of the Koszul-Tate differential to vanish identically under a
proper `rotation' of the antighost number one antighosts such that the total
number of physical degrees of freedom does not vary. The irreducible
Hamiltonian analysis of the initial quadratic action presents the desirable
feature that it induces a corresponding irreducible Lagrangian version,
which, in turn, is the most natural framework for investigating higher-order
interacting Lagrangian gauge theories with topological coupling. Finally, we
remark that our analysis covers the free case in the limit $M\rightarrow 0$.

\end{document}